# Temperature-dependent photoluminescence: A theoretical study

## M. Kurtulik,[1] A. Manor,[1] R. Weill,[2] and C. Rotschild[1,2,*]

[1]Russell Berrie Nanotechnology Institute, Technion – Israel Institute of Technology, Haifa 3200003, Israel
[2]Department of Mechanical Engineering, Technion – Israel Institute of Technology, Haifa 3200003, Israel
*Corresponding author: carmelr@technion.ac.il

**Photoluminescence (PL) is a light–matter quantum interaction associated with the chemical potential µ of light formulated by the Generalized Planck's law. Without knowing the inherent temperature dependence µ(T), the Generalized Planck's law is insufficient in order to characterize PL(T). Recent experiments showed that PL at a critical temperature abruptly shifts from a conserved rate, accompanied by a blue-shift, to thermal emission. Here, we theoretically study temperature-dependent PL by including phononic interactions in a detailed balance analysis. We show that in a three-level system, both µ and T are defined in the case of fast thermalization. Our solution validates recent experiments and predicts new features, including an inherent relation between emissivity and external quantum efficiency of a system, a universal point defined by the pump and the temperature where the emission rate is fixed to any material, a new phonon induced quenching mechanism, and thermalization of the photon spectrum. Our high-temperature luminescence solution is relevant to and important for all photonic fields where the temperature is dominant.**

Photoluminescence (PL) conventionally involves the absorption of high-energy phonons followed by an emission of red-shifted low-energy photons. PL, first studied by Stokes [1], has been extensively researched by many others [2, 3, 4, 5, 6, 7]. Due to the complexity of many-body interaction, PL(T) on the microscopic scale is challenging to formulate; thermodynamics, however, allows it to be statistical analyzed [8, 9, 10, 11]. Treating light as ideal gas particles means that PL can be described using the usual thermodynamic variables such as temperature and chemical potential [12, 13, 14]. PL at elevated temperatures exhibits properties that are very different from and even counter-intuitive to thermal emission. Figure 1 shows typical temperature-dependent PL and thermal photon rates (counts per second) of Nd$^{3+}$ ions in glass excited by a 532nm CW laser (Fig. 1a), with their corresponding power spectrum (Fig. 1b) under constant incident power taken from [15].

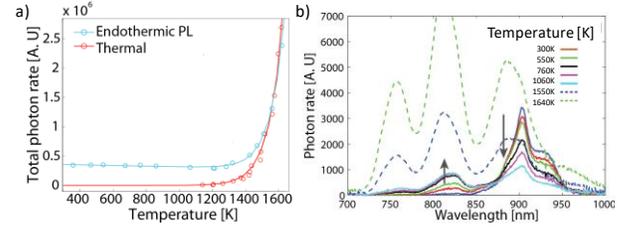

Figure 1. a) Nd$^{3+}$ temperature-dependent PL (blue) and thermal (red) photon rates showing the critical temperature, $T_c$. b) Emission spectrum vs. temperatures showing rate-conservation and blue-shift below $T_c$ (solid lines), followed by a transition to the thermal regime above $T_c$ (dashed lines).

As can be seen, the PL rate (blue line in Fig. 1a) is conserved until a critical temperature, $T_c$, is reached. Above $T_c$, as the total rate increases, the emission converges to the thermal emission (red line). Looking at the spectrum (Fig. 1b), at temperatures lower than $T_c$ (solid lines), low-energy photons exhibit a blue-shift towards high energy photons. This blue-shift under a constant rate results in a reduction of low energy photons with increasing the temperature and compensation in the rate of high-energy photons. It is conventionally used in optical refrigeration [16, 17]. This phenomenon contrast sharply with Planck's radiation, where the rate monotonically increases with temperature at any wavelength. As can be seen in Figure 1b, above $T_c$ (dashed lines), the emission becomes thermal, which is monotonically increasing at any wavelength. To the best of our knowledge, there is no theoretical model supporting such behavior for arbitrary quantum efficiency (QE) or external quantum efficiency (EQE).

To explain the experimental data, we begin with the description of any light source given by the Generalized Planck's formula ascribing temperature and chemical potential to PL emission [13, 14]:

$$L(h\nu, T, \mu) = \varepsilon(h\nu) \cdot \frac{2\nu^2}{c^3} \frac{1}{e^{\frac{h\nu-\mu}{k_B T}}-1} \qquad (1)$$

where $L$ is the spectral radiance (having units of watt per frequency, per solid angle, and per unit area), $T$ is the temperature, $\varepsilon$ is the emissivity, $h\nu$ is the photon energy, $K_B$ is Boltzmann's constant, and $\mu$ is the chemical potential, which is the Gibbs free energy per emitted photon or the gap that is opened between the quasi-fermi-levels under excitation. The described emission is at a specific frequency band, where the chemical potential is constant and can be defined by its brightness temperature, $T_{Brightness}$, which is the black body temperature having the same radiance at the same frequency band. According to Equation (1), $T_{Brightness} = T/(1 - \mu/h\nu)$. As can be seen, for any $T_{Brightness}$, $\mu$ and $T$ are not uniquely defined and for $\mu = 0$, the emission becomes thermal; $T_{Brightness} = T$.

The QE is defined by the ratio between the emitted photon rate and the absorbed pumped photon rate for a material at low temperatures when thermal excitation is negligible and reflects the competition between radiative and nonradiative relaxations. External-QE (EQE) is defined by the ratio between the incoming and the outgoing photons from a cavity together with the material and, in addition to the material properties, depends on the cavity. Let us first consider the simple case depicted in Figure 2a. There a PL body is represented by two energy levels with some absorptivity $\alpha$ and an emissivity $\varepsilon$, satisfying Kirchhoff law $\alpha = \varepsilon$, having some QE located inside an optical cavity, in thermal contact with a heat reservoir at temperature $T$ and coupled to an optical pump source with brightness temperature $T_p$ and coupling coefficient $\Gamma_{pump}$ which excites the PL body above thermal excitation, $\mu > 0$. To understand the quasi-equilibrium temperature evolution, consider the temperature-dependent emission rate at a specific frequency and emission coupling coefficient $\Gamma_{out}$ for the two **approximating** cases of zero-EQE and unity-EQE. In the former case, the nonradiative relaxation rate is dominant, resulting in thermal emission with emissivity $\varepsilon$, regardless of optical pump excitation, and $\mu = 0$ (Fig. 2b, black line). The latter case describes the absence of nonradiative channels. In other words, thermal excitation cannot promote electrons above the bandgap, resulting in a conservation of the emitted photons as the temperature changes (Fig. 1b, dashed red line). In the symmetric case where $\Gamma_{pump} = \Gamma_{out}$, these two lines intersect at a critical point where the thermal emission rate equals the absorbed pump rate, reflecting equilibrium between the optical pump source and the heat reservoir $T_p = T$, which suggests that there is zero Carnot efficiency between two energy sources (reservoirs) when their temperatures are the same [12]. This implies that the resulting PL spectrum converges to the thermal emission; $\mu = 0$, with $T_{Brightness} = T$. Moreover, the emission of any EQE restricts it to in-between these two lines (Fig. 1b, blue area) and at the critical point becomes a universal point. In the non-symmetric case where $\Gamma_{pump} < \Gamma_{out}$, the critical point accurs at lower temperature.

We note that for the case of unity-EQE, the case of the absence of radiative channels results in absorptivity and emissivity approaching zero, and the material is transparent.

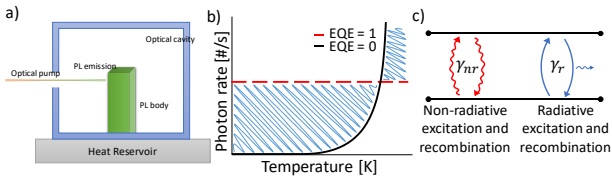

Figure 2. a) A PL body in contact with a heat reservoir at temperature **T** and an optical pump at brightness temperature $T_p$. b) The photon rate of a PL body with EQE=1 (red line) and EQE=0 (black line). The blue area shows any other EQE limits on the emission rate. c) A general two-level system having radiative and nonradiative interactions.

Figure 2c shows the different mechanisms involved in the detailed balance of the rates upon absorption of photons where radiative $\gamma_r$ and nonradiative $\gamma_{nr}$ rates are competing through spontaneous and stimulated processes [16]. $\gamma_{nr}$ also allows for thermal excitation, which—at thermal equilibrium—balances all nonradiative processes. This mechanism reveals the origin of the universal critical point, as depicted in Figure 2b. When the pump's brightness temperature is equal to the reservoir temperature $T_p = T$, the system is at thermal equilibrium; thus, thermal excitation cancels nonradiative recombination. The cancelation of nonradiative processes results in emission that appears as having both unity-EQE and thermal emission.

In the following, we extend this intuitive picture to a three-level system in a cavity and thereafter show the generality of the solution to any system with or without a cavity.

By following Siegman [16], we study the case of a three-level system in a cavity (describing, for example, the 750nm–900nm emission lines of Nd$^{+3}$ [19,20] depicted in Fig. 1). Such a detailed balance considers only photonic electronic and phononic transitions and omits other processes such as defects, which may cause temperature-dependent quenching. As such, this model describes the upper limit of temperature-dependent luminescence. Nevertheless, any additional factors can be embedded in the radiative and nonradiative rates for a specific solution. Here we assume the radiative and nonradiative rates to be temperature independent. Additional temperature-dependent values, such as the bandgap reduction with temperature rise in semiconductors, can be implemented in the model.

Figure 3a shows the considered energy levels having a ground state and a broad excited level consisting of two closely spaced levels, with very fast nonradiative thermalization between them. This ensures a Boltzmann distribution of excited states, $n_2$ and $n_3$ [121]. Such a system is described by:

$$\frac{dn_2}{dt} = (n_1 - n_2)B_{r12}n_{ph12} - n_2\gamma_r + (n_1 - n_2)B_{nr12}n_{pn12} - n_2\gamma_{nr} + (n_3 - n_2)B_{nr23}n_{pn23} + n_3\gamma_{nr23} \quad (2a)$$

$$\frac{dn_3}{dt} = (n_1 - n_3)B_{r13}n_{ph13} - n_3\gamma_r + (n_1 - n_3)B_{nr13}n_{pn13} - n_3\gamma_{nr} - (n_3 - n_2)B_{nr23}n_{pn23} - n_3\gamma_{nr23} \quad (2b)$$

$$4\pi \cdot \Delta\nu \cdot \frac{dn_{ph12}}{dt} = -n_{ph12}\Gamma_{12}c - (n_1 - n_2)B_{r12}n_{ph12} + n_2\gamma_r \quad (2c)$$

$$4\pi \cdot \Delta\nu \cdot \frac{dn_{ph13}}{dt} = n_{pump}\Gamma_p c - n_{ph13}\Gamma_{13}c - (n_1 - n_3)B_{r13}n_{ph13} + n_3\gamma_r \quad (2d)$$

where $n_1, n_2, n_3$ are the electron population densities of the ground and excited states, respectively; $c$ is the speed of light; $\gamma_r, \gamma_{nr}$ are radiative and nonradiative spontaneous rates, from both upper levels to the ground state, with units of [1/s], respectively; $\gamma_{nr23} \gg \gamma_r, \gamma_{nr}$ is the nonradiative rate between excited states $n_2$ and $n_3$; $\Gamma_p, \Gamma_{12}, \Gamma_{13}$ are coupling rates in and out of the cavity, respectively; $n_{ph}$ is the radiation field density inside the cavity, having units of $\left[\frac{\#}{\Delta\nu \cdot Sr \cdot m^3}\right]$; $n_{pump} = DoS_{ph}\left[\exp\left(\frac{E_g}{kT_p}\right) - 1\right]^{-1}$ is the optical pump; and $B_r = \gamma_r/DoS_{ph}$, and $B_{nr} = \gamma_{nr}/DoS_{pn}$ are the Einstein coefficients [8]. Under fast thermalization, phonons obey the equilibrium distribution and $n_{pn}$ is given by $n_{pn} = DoS_{pn}\left[\exp\left(\frac{E_g}{kT_p}\right) - 1\right]^{-1}$ [12]. In this formalism, $DoS_{pn}$ and $DoS_{pn}$ are the corresponding densities of states (DoS) for the phonons and photons of the material and of the cavity. The solutions of equations (2a–2d) for the spectrally integrated photon rate (total radiative-transitions to the ground state), for fixed output coupling rates $\Gamma_{12} = \Gamma_{13} = \Gamma$, under $T_p = 1000K$ and various EQEs (0, 0.5, 1), are depicted in Figure 3b. The figure shows the quasi-rate-conservation accompanied by the blue-shift of the spectrum (inset) below the universal point $T_c$, defined by the crossing with the pump rate, $n_{pump}\Gamma_p$ (red line).

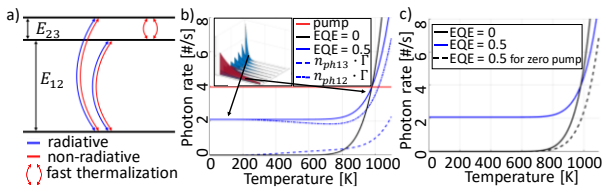

Figure 3. a) A three-level system with fast thermalization between energy levels $n_2$ and $n_3$. b) For EQE=0.5 and pump brightness temperature $T_p = 1000K$, insets show a blue-shift of the spectrum with temperature. c) Above the critical temperature $T_c < T$, the PL emission (blue line) is bounded by a black body (EQE=0, solid black line) and the thermal emission (dashed black line).

The inset depiction of the blue shift was extrapolated from the discrete energy level solution, shown as the breakdown of the total emission rate (solid blue line) into its two different energies: $E_{12}$ (dot-dashed blue line) and $E_{13} = E_{12} + E_{23}$ (dashed blue line). The ratio of these individual emissions is given by the Boltzmann distribution for our case of $\gamma_{nr23} \gg \gamma_r, \gamma_{nr}$ [21]. With a further temperature increase, the photon rate rises to the universal point, where all EQE intersect. This general solution, is as far as we know, the first theoretical explanation for the experimentally observed transition from rate-conservation accompanied by a blue-shift (reduction in low-energy-photons) to thermal emission, where the photon rate increases at any wavelength. Furthermore, figure 3c shows the emitted photon rate from a PL body having 50% EQE (solid blue line), and the curve for EQE=0 times the output coupling rate, $\Gamma$ (solid black line). Setting EQE=0 describes the thermal emission case and is invariant to the pump rate. In addition, the thermal emission for the 50% EQE, and a cavity with coupling-rate $\Gamma$, are considered when setting the pump to zero $n_{pump} = 0$ (depicted by the dashed black line), which results in a reduced thermal emission with respect to the zero-EQE thermal curve. We note that all thermal and non-thermal emission curves collapse to the black body line in the case of a closed cavity ($\Gamma = 0$). The thermal emission for EQE>0 is reduced compared to EQE=0 due to a lower value of $\gamma_{nr}$ compared to $\gamma_r$. As evident, the emission beyond the universal point is restricted to remaining between the zero-EQE and the thermal curve for the same EQE. The asymptotic behavior of the PL and thermal emissions is also evident in the experimental data (Fig. 1a). In the supplementary material, we show how opening the cavity (a lower Q-factor or increasing $\Gamma$) shifts the asymptotic region closer to the universal point.

The **ratio** between the thermal curve for a specific material (EQE>0) and the zero-EQE emission curves is a temperature-independent and EQE-dependent constant we name $\varepsilon_{EQE}(\gamma_r, \gamma_{nr})$.

Figure 4a depicts a linear relation between this constant and EQE. Approaching unity-EQE demands $\varepsilon_{EQE}(\gamma_r, \gamma_{nr}) \approx 0$, whereas approaching zero-EQE demands $\varepsilon_{EQE}(\gamma_r, \gamma_{nr}) \approx 1$. This is, as far as we know, the first indication of an inherent dependency between EQE and emissivity. In general, the emissivity can be written as the multiplication of the emissivity of zero-EQE, which is only DoS dependent, and the emissivity factor, which is EQE dependent:

$$\varepsilon_{tot} = \varepsilon_{zero-EQE}(DoS) \cdot \varepsilon_{EQE}(\gamma_r, \gamma_{nr})$$

Figure 4b shows the relation between pairs of $\gamma_r$ and $\gamma_{nr}$ and EQE. As the coupling rate $\Gamma$ approaching zero, EQE approaches 0 (and the emissivity approaches unity). In our model, $\gamma_r$ and $\gamma_{nr}$ are not $\Gamma$ dependant. This is the case when cavity dimensions are much larger than emission wavelength and the Purcell factor converges to unity [22].

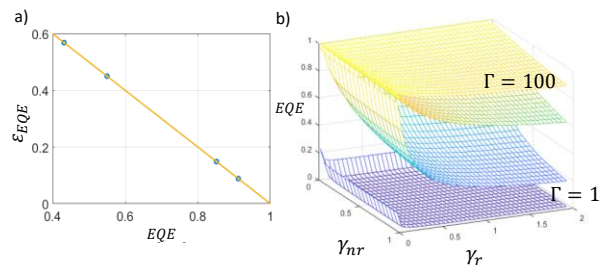

Figure 4. a) Linear relation between the open-cavity emissivity $\varepsilon_{OC}$ and EQE. b) QE for various radiative and nonradiative rates vs. the coupling rate $\Gamma$.

Thus far, we considered constant DoS with wavelength. When describing the emissivity line-shape, we set different DoS for both excited levels. Figure 5 shows the solution for such a case, which fits the experimental results depicted on Figure 1 better.

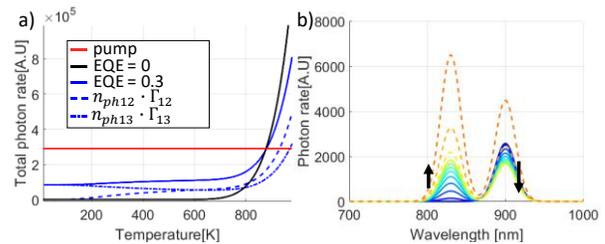

Figure 5. Different DoS per excited level. a) Total photon rate for EQE=0.3. b) The spectrum exhibits blue-shift and transition to thermal rate.

Interestingly, our model goes beyond the specific experiment described in Figure 1 and predicts a new temperature induced quenching. Figure 6a depicts a system having a broad ground state represented by two energetically closed low energy levels. Figure 6b shows a decrease in the photon rate below the critical temperature due to a stimulated nonradiative interaction. At low temperatures when phononic excitation is negligible, $n_2$ is empty, and the spontaneous radiative and nonradiative recombination from the $n_3$-level becomes involved. Increasing the temperature slightly results in a faster depletion in $n_3$ due to stimulated nonradiative recombination, which surmounts the weak thermal excitation. A further temperature rise leads again to an increase in the $n_2$ population, and an increase in the $n_3$ population, due to stimulated absorption, results in enhanced emission. This behavior is also seen in our experimental result [15]. Finally, our model also supports recent experiments and theory claiming the thermalization of the PL spectrum and emissivity approaching unity when closing the cavity (minimizing Γ). It was shown that in such a case, the spectrum evolves into a Boltzmann distribution [23, 24] (see supplementary material for a detailed explanation).

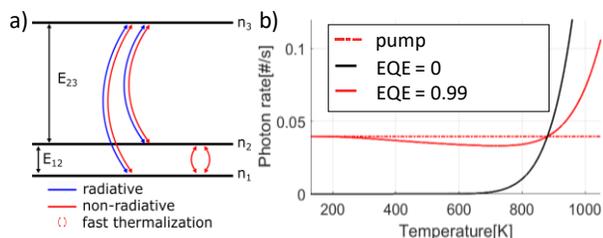

Figure 6. a) The three-level system with a broad ground state, where the energy levels $n_1$ and $n_2$ are non-radiatively coupled. b) A simulation of the given system with a sudden drop in emission due to stimulated nonradiative recombination and the critical point for various EQEs.

In summary, we developed the model for temperature-dependent luminescence using a detailed balanced formalism at high temperatures, where the thermal excitation is comparable to the photonic excitation. Our results support the experimental observations of photoluminescence at elevated temperatures, exhibiting a blue-shift of the spectrum while the photon rate is conserved and the transition to thermal emission. We also show the existence of a universal point where the emission rate of any EQE, under a fixed incident pump and temperature, is set. More generally, our model is the first to show an inherent dependency between EQE and emissivity. Our model can be important in lighting and energy harvesting systems as well as to any field of radiation where the evaluation of the limit of radiation is critical.

**Funding.** European Union's Seventh Framework Program (H2020/2014-2020]) under grant agreement n° 638133-ERC-ThforPV

**Disclosures.** The authors declare no conflicts of interest.

See Supplement 1 for supporting content.